# Microwave Power Absorption to High Energy Electrons in the ECR Ion Thruster

Giulio Coral, Ryudo Tsukizaki, Kazutaka Nishiyama, and Hitoshi Kuninaka

**Abstract**

The microwave power absorption efficiency of the µ10 ECR ion thruster, utilized in the Japanese asteroid explorers Hayabusa and Hayabusa2, is investigated in order to allow performance measurement and provide information for its improvement. A model detailing the local electron behavior in a real ECR plasma discharge, based the magnetic field characteristics, is presented. Three methods to evaluate the microwave power absorption efficiency are proposed: an estimation based on the chamber geometry and magnetic field characteristics, a measuremen based on performance parameters and a measurement performed with Langmuir probes. The equations used for each method are analytically derived. The local electron behavior model is confirmed with a Langmuir probe experiment. Measurement of the microwave power absorption efficiency is performed with the two independent methods proposed. Results from the two experiments show good agreement with each other and with the theory. Finally, a diffusion model explaining the different electron temperature distributions observed in the chamber is proposed. The model and experiments clarify the physics behind previously observed performance variations and give valuable hints for future chamber improvement.

1. Introduction

Electric propulsion is a general classification that includes all the space engines that utilize electricity to increase the exhaust velocity (to increase the specific impulse). While their thrust is lower than chemical rocket engines, they require a considerably lower amount of propellant to accomplish the same space mission, allowing deep space exploration and prolonged satellite lifetime.

Among them, ion thrusters are those offering the highest specific impulse and reliability, making them the most suitable for interplanetary robotic missions. These engines produce plasma in a discharge chamber, and accelerate it through high-voltage grids. The discharge chamber layout has taken different paths for the main ion thruster developing countries, with the US using DC discharges, Europe using RF waves and Japan adopting microwave electron cyclotron resonance (ECR) heating.

The µ10 ECR ion thruster and its ECR neutralizer were developed to tackle the potential lifetime issues of DC thrusters, especially the use of hollow cathodes used for primary electron generation and ion beam neutralization [1-2]. Thanks to the high reliability achieved, µ10 was capable of powering Hayabusa, the first asteroid sample return mission (towards the Itokawa asteroid), succesfully completed in June 2010 [3]. An upgraded version of the thruster is presently utilized in the Hayabusa2 probe [4,13], launched in December 2014 and targeting the Ryugu asteroid, as it will be in the future for the DESTINY+ mission.

The physics behind plasma formation in the thruster, especially with regard to performance improvements achieved in the past [5], have not been fully understood yet. This paper will build upon the 0-dimensional model for ECR thruster, based on our previous work [6], and develop a 2-dimensional (the thruster has rotational symmetry) theoretical framework that will prove useful in the engineering process of µ10. The 0-dimensional model for ECR thruster includes several parameters, such as the microwave transmission rate, the baseline plasma ion energy cost and the ion current ratio, which are easily measured. The input microwave power will be distributed to the reflection power $P_R$ from the discharge chamber, the wall loss $P_w$ in the chamber, the heating to unconfined electrons and the heating to confined electrons, of which only the final component contributes to generate ion. The microwave power absorption efficiency α, which is defined as the ratio of microwave power transferred to high energy electrons confined in the mirror magnetic field (which in turn ionize neutrals) to the net microwave power input, hasn't been investigated yet. This paper aims to develop the theory about α and to quantify it.

## 2. Methods
### 2.1. The µ10 ECR Ion Thruster

µ10 is an ion thruster with 10cm diameter high-voltage grid system. It utilizes an ECR discharge to ionize xenon gas. Propellant enters the chamber through multiple injectors, while 4.25GHz microwave power input comes from an antenna located in the waveguide (far from the plasma discharge to increase its lifetime). Permanent magnets, with 0.4T magnetic field strength at their surface, are located in the discharge chamber, and form the magnetic mirror where the ECR discharge occurs. Xenon ions produced in the discharge are extracted through a high-voltage ion acceleration grid system, that ejects them at high velocities [14]. The thruster schematic is shown in Figure 1.

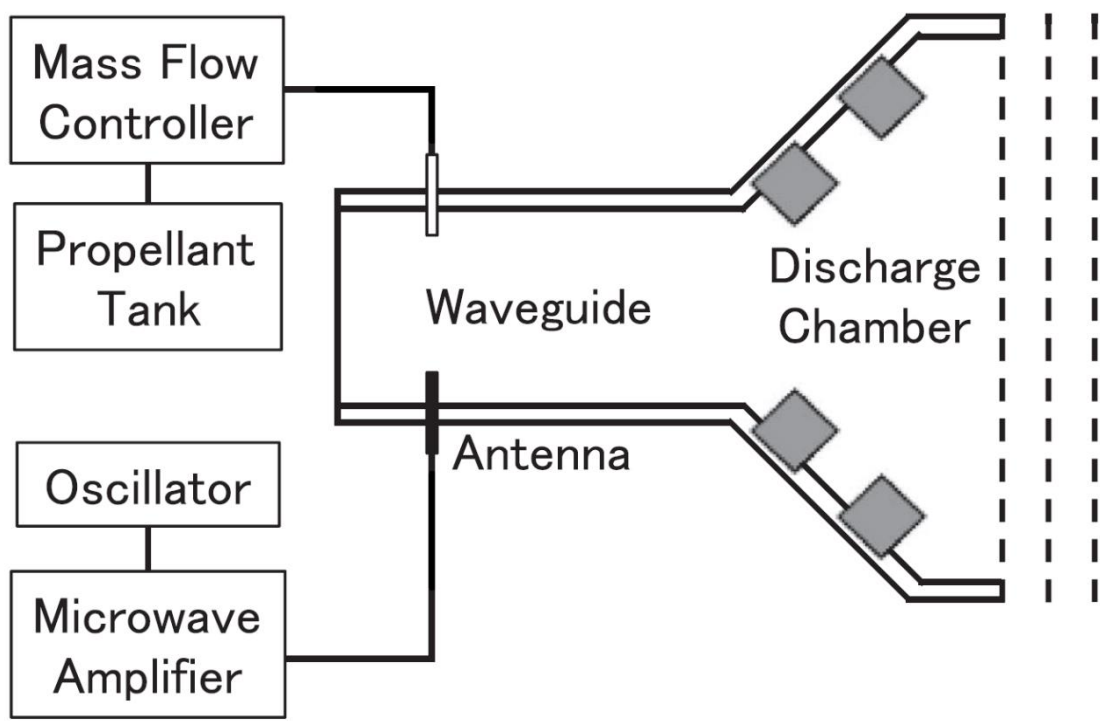

*Figure 1: µ10 ECR Ion Thruster Schematic*

The µ10 ino thruster and its neutralizer have already achieved, through a series of improvements, fully satisfactory levels, surpassing the mark of 10,000h continuous operation. µ10 has been improved since the first flight version, and its 40% total efficiency qualifies it as one of the best performing sub-500W electric propulsion systems.

However, the plasma physics behind its improvement have not been revealed yet. To understand these processes, a 0-dimensional ion production model for ECR ion thrusters has been developed in 2007 [6], following the outline of its equivalent DC thruster model [7]. Equivalent models have been developed for RF ion thrusters [12], and are supporting the development of such systems. The 0-dimensional ECR model pointed out the necessity of investigating the microwave power absorption efficiency, especially focusing on the operational and design parameters affecting it.

### 2.2. 2-Dimensional Model for the ECR Discharge Chamber

ECR occurs when the cyclotron gyration frequency of an electron in a magnetic field corresponds to the frequency of the microwave power input from a given source. In a real geometry, electrons will go through an isomagnetic surface, gaining an energy $\varepsilon_{ECR}=eV_{ECR}$, where the ECR voltage $V_{ECR}$ is defined as [8]:

$$V_{ECR} = \frac{\pi E_{ECR}^2}{v_{//} \left|\frac{\partial B}{\partial s}\right|_{ECR}}$$

(1)

As a single passage is not sufficient to bring the electrons to the energies required for ionization, ECR discharge chamber normally involve magnetic mirror confinement. In closed magnetic mirrors as in Fig. 2, electrons will be trapped or lost to the magnets depending on the pitch angle θ between their velocity and the magnetic lines. This will be represented by the loss cone factor (or mirror ratio) R:

$$R = \frac{1}{(\sin\theta_{mag})^2} = \frac{B_{mag}}{B_{min}}$$

(2)

Where $B_{mag}$ and $B_{min}$ are respectively the maximum (found on the magnet surface) and minimum magnetic field on the magnetic tube. Similarly, we can define $\theta_{ECR}$ as in equation 2 but using $B_{ECR}/B_{min}$: electrons with a pitch angle between these two are confined and pass through the ECR region multiple times while mirrored in the magnetic mirror. It must be pointed out that this description is valid, for a single electron, as long as it does not collide with other particles, since collisions will change its momentum and pitch angle. However, considering the whole population of electrons, this effect will be balanced by equal and opposite momentum transfers, meaning that our definition will remain valid on average. A schematic representation of ECR heating in a magnetic mirror is shown in Figure 2.

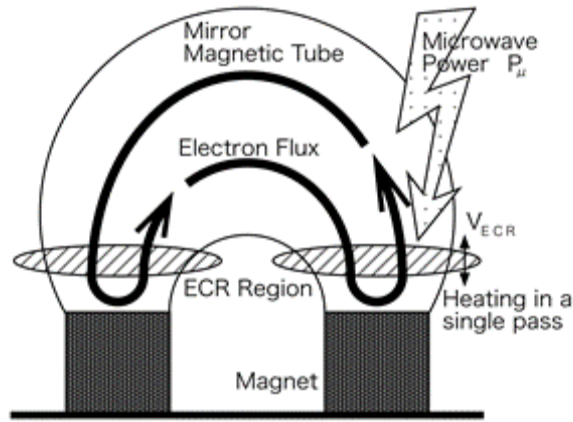

*Figure 2: Schematic of the ECR Heating in a Magnetic Mirror*

Due to further constraints, such as electron drifts and beam extraction, the configuration chosen for µ10 and other ECR ion thrusters involves two opposite polarity magnetic rings, forming a curved magnetic mirror. It can be noticed how this configuration has magnetic lines not crossing the ECR and open-ended magnetic bottles. We can separate the chamber in three regions, as shown in Figure 3, which we will define as:

- Region 1: closed magnetic mirrors crossing the ECR isomagnetic. Electrons cross it multiple times gaining $E_{ECR}$ at each passage.
- Region 2: closed magnetic mirrors not crossing the ECR isomagnetic. No electron heating occurs.
- Region 3: open magnetic bottles crossing the ECR isomagnetic. Electron can gain up to $2E_{ECR}$ before being lost to the system walls.

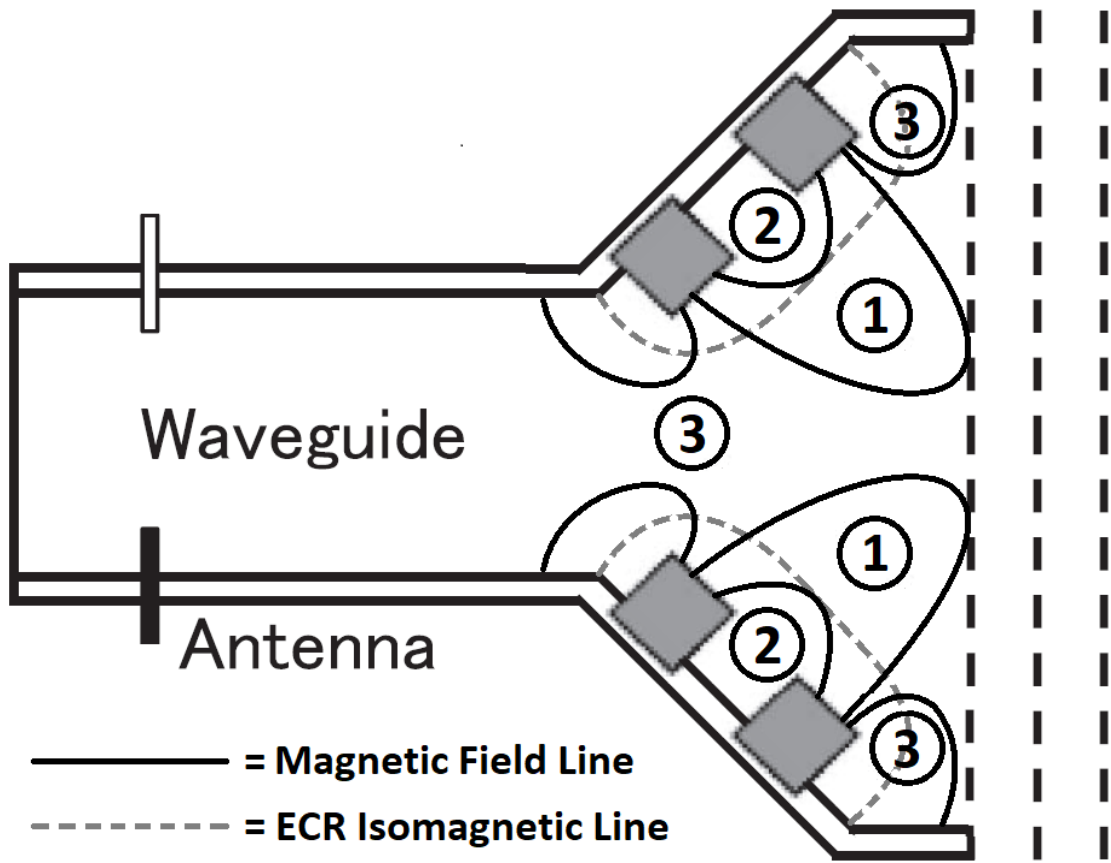

*Figure 3: Regions Subdivision in the Ion Thruster Chamber*

Following these considerations, it is clear that the plasma density in Region 1 is expected to exceed those of Region 2 and 3.

### 2.3. The Microwave Power Absorption Efficiency Equations
#### 2.3.1. Estimation of α from the Microwave Power Distribution Based on Chamber Geometry

The simplest method we propose to estimate the microwave power absorption efficiency is based on the analysis of the discharge chamber geometry.

Microwave power coming from the antenna in the waveguide is absorbed at the ECR isomagnetic surface, shown in Fig. 3, where microwave frequency and electron cyclotron frequency are equal [9]. Based on the three regions model, this will lead to:

- Region 1: power is effectively transferred to electrons.
- Region 2: no power is absorbed, as electrons do not cross the ECR region.

- Region 3: power is dissipated by the electrons.

Furthermore, electrons in Region 1 can be divided in three different groups, depending on their pitch angle:

- Region 1A: electrons with a pitch angle larger than $\theta_{ECR}$ never reach the ECR region, hence they are not associated with the energy transfer.
- Region 1B: electrons with a pitch angle smaller than $\theta_{ECR}$ and larger than $\theta_{mag}$ are effectively heated by microwave due to multiple passes through the ECR region.
- Region 1C: electrons with a pitch angle smaller than $\theta_{mag}$ absorb the microwave power and dissipate it into the wall due to the open magnetic bottles.

$\theta_{ECR}$ and $\theta_{mag}$ are calculated from equation 2 using $B_{ECR}/B_{min}$ and $B_{mag}/B_{min}$ respectively. The ECR magnetic field intensity $B_{ECR}$ is 0.15T associated with 4.25GHz microwave. In other words, the microwave power is transferred to Regions 1B, 1C and 3, but the only Region 1B contribute to generate ions. The ratio of the microwave power transferred to Region 1B against those to Regions 1B, 1C and 3 is equivalent to α. Taking into account the power loss $P_W$ (direct wall heating) and the power reflection $P_R$ from the chamber, we can formulate the microwave power absorption efficiency equation as:

$$\alpha = \left(1 - \frac{P_W + P_R}{P_\mu}\right) \frac{A_{R1}}{A_{R1} + r A_{R3}} \left(1 - \frac{1 - \cos\theta_{mag}}{1 - \cos\theta_{ECR}}\right)$$

(3)

In which $A_{R1}$ and $A_{R3}$ are the surface areas of the ECR isomagnetic plane in each region ($A_{R2}$, based on the considerations we made, is zero). The electrons in Region 1 must absorb microwave power more effectively, as its density is much larger than that of Region 3 (as discussed in section 3), so we introduce the density ratio r.

Using this equation, we can estimate the microwave power absorption efficiency exclusively from the thruster design parameters. With $\theta_{ECR}=54°$, $\theta_{mag}=28°$, $A_{R1}=74cm^2$, $A_{R3}=68cm^2$, r=0.14 and neglecting $P_W$ and $P_R$, this leads to α=0.52. This estimation, which doesn't take into account losses and other features of real operations, will be kept as a guideline upper limit.

### 2.3.2. Estimation of α from the Ion Production Performance

The second method we will use to measure the microwave power absorption efficiency, in order to strengthen our considerations with experimental results, is based on the equation relating α with the ion production cost $C_i$, for which the derivation can be found in the 0-dimensional ion production model [6]:

$$\alpha = \frac{\varepsilon_P^*}{C_i \ f_s \left(1 - e^{-C_0 \dot{m}(1-\eta)}\right)}$$

(4)

In which $\varepsilon_p^*$ is the baseline plasma ion energy cost, $f_s$ is the ion current ratio (≈0.4) and $C_0$ is the primary electron utilization factor. Equation 4 is represented in Figure 4 with the α-$C_i$ chart for selected values of the propellant utilization efficiency η.

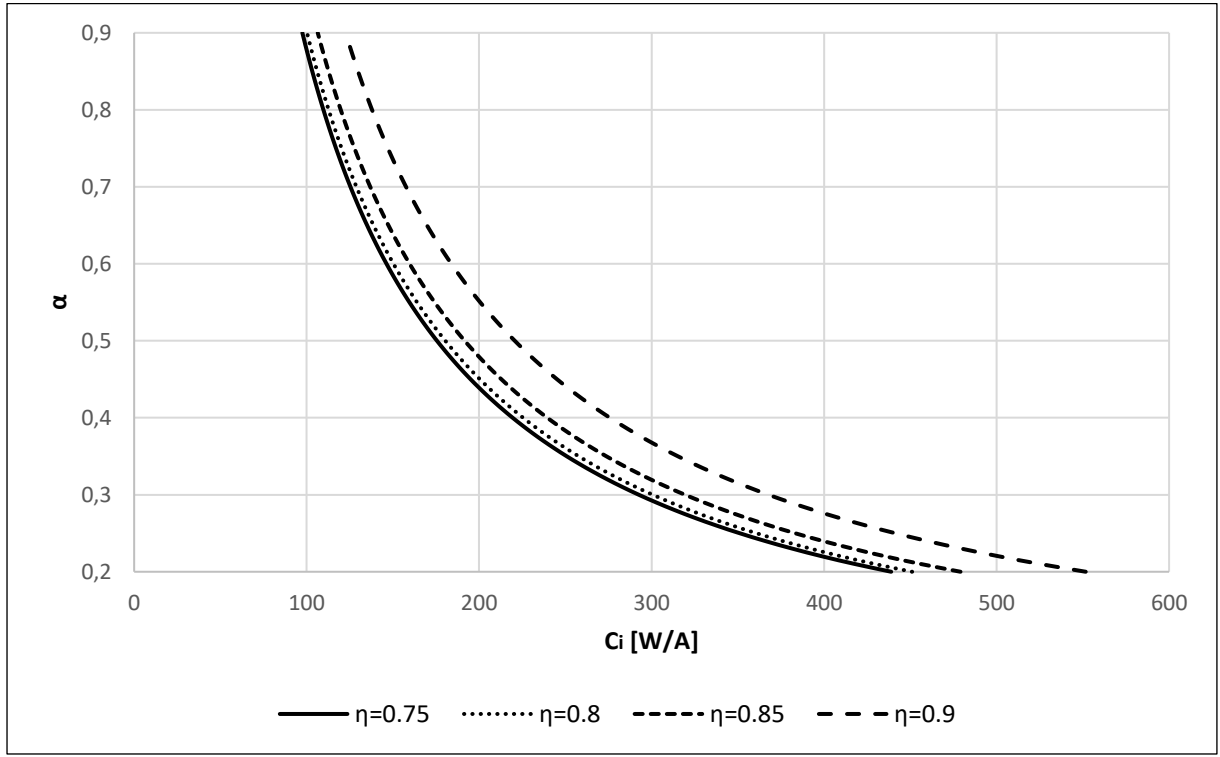

*Figure 4: Predicted Values of Microwave Power Absorption Efficiency for Different Propellant Efficiencies*

As $C_0\dot{m} \approx 10$ [6], we can simplify (except for $\eta \rightarrow 1$) equation 4 as:

$$\alpha \approx \frac{\varepsilon_P^*}{C_i\, f_s}$$

(5)

This represent a horizontal asymptote in the η-$C_i$ chart, as shown in Figure 5, for cases in which the baseline plasma ion energy cost $\varepsilon_p$* is independent on the neutral density, such as DC ion thrusters [7]. For ECR ion thrusters $\varepsilon_p$*, as electrons are heated progressively, a higher neutral density will lead to a lower electron energy: hence, at high values of ṁ (low η), $C_i$ is expected to increase. Taking $\varepsilon_p$* dependent and α independent on η, we will be able to evaluate it without the disturbance caused by probes (both η and $C_i$ are calculated from input and performance parameters).

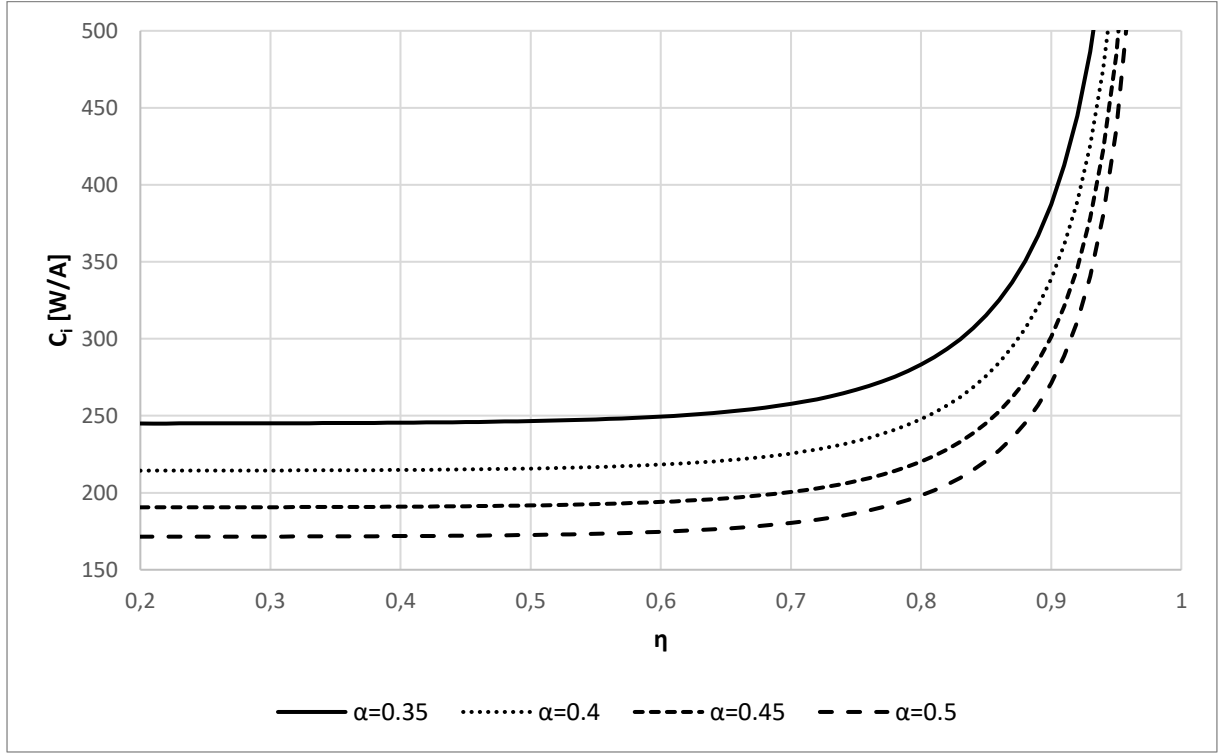

*Figure 5: η-$C_i$ Chart with Horizontal Asymptote for Constant $\varepsilon_p$\**

### 2.3.3. Estimation of α from the Electron Heating Process

The microwave power absorption efficiency α is defined as [6]:

$$\alpha = \frac{4V_{ECR}I_e}{P_\mu}$$

(6)

Where $P_\mu$ is the microwave power input, $I_e$ is the electron current across the ECR isomagnetic surface and the 4 factor is introduced as an electron will cross the ECR 4 times in a single cycle.

We will seek now to obtain a more informative equation for α, showing which design parameters affect it and allowing its measurement with Langmuir probes. We start by representing electrons at a given temperature in the velocity space as spheres of radius v (dependent on the electron temperature $T_e$), as shown in Figure 6.

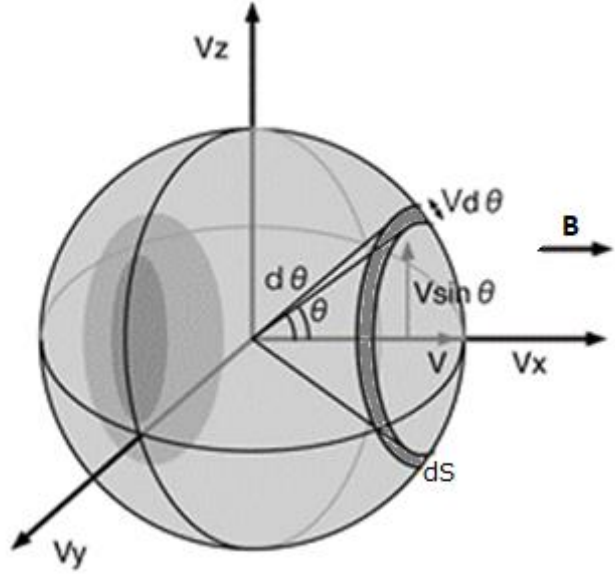

*Figure 6: Sphere in the Velocity Space*

Multiple spheres will define the distribution function F(v). In the surface of a sphere, we will find a number of electrons as:

$$dn = F(v)dv$$

(7)

Electrons in the truncated spherical sector dS will have a velocity parallel to B:

$$v_{//} = v\cos\theta$$

(8)

The area of this sector will be:

$$dS = 2\pi v \sin\theta \, v \, d\theta$$

(9)

Hence, we will find here a number of electrons:

$$dndS = F(v)dv \frac{2\pi v \sin\theta \, v \, d\theta}{4\pi v^2}$$

(10)

From equations 7, 8 and 10 we can determine the current flowing from this sector to the ECR region:

$$dI = e\, v \cos\theta \; A_{min}\, dn\, dS = \frac{1}{2} e\, v\, F(v)\, A_{min} \cos\theta \sin\theta \, dv\, d\theta$$

(11)

Where $A_{min}$ is the area of the magnetic mirror at the minimum magnetic field. By substituting equations 1 and 11 in equation 6, the microwave power absorption efficiency is rewritten as follows:

$$\alpha = \frac{2\pi E_{ECR}^2}{v\cos\theta\, P_\mu \left|\frac{\partial B}{\partial s}\right|_{ECR}} \iint dI = \frac{2eA_{min}\pi E_{ECR}^2}{P_\mu \left|\frac{\partial B}{\partial s}\right|_{ECR}} \int_0^{+\infty} F(v)dv \int_{\theta_{mag}}^{\theta_{ECR}} \sin\theta\, d\theta$$

(12)

In which $\theta_{mag}$ and $\theta_{ECR}$ are calculated from equation 2, using $B_{mag}/B_{min}$ and $B_{ECR}/B_{min}$ respectively.

The first integral is easily solved as:

$$\int_0^{+\infty} F(v)dv = n_e$$

(13)

While the integration boundaries of the second allow us to consider only the electrons crossing the ECR and not being lost to the walls. Hence, we get:

$$\alpha = \frac{2eA_{min}\pi E_{ECR}^2}{P_\mu \left|\frac{\partial B}{\partial s}\right|_{ECR}} n_e |-\cos\theta|_{\theta_{mag}}^{\theta_{ECR}}$$

(14)

This formulation, while less synthetic, offers valuable information about the operational and design parameters affecting microwave power absorption. From an experimental point of view:

- $A_{min}$, $\theta_{ECR}$, $\theta_{mag}$ and $\partial B/\partial s$ are determined by the ECR ion source geometry
- $P_\mu$ is the microwave power input
- $E_{ECR}$ is measured by electro-optical probes [10]
- $n_e$ is measured by Langmuir probes

**3. Results**

**3.1. Measurement of the Plasma Properties**

Our first experiment aims to observe the plasma properties of Regions 1, 2 and 3, as considerations made in section 2.2 suggest there should be differences between them. Three Langmuir probes are placed on the downstream magnet's surface (to minimize the Langmuir probes disturbance to the microwave electric field) in correspondence of the three regions. The setup is shown in Figure 7 and 8, with the experimental conditions reported in table 1.

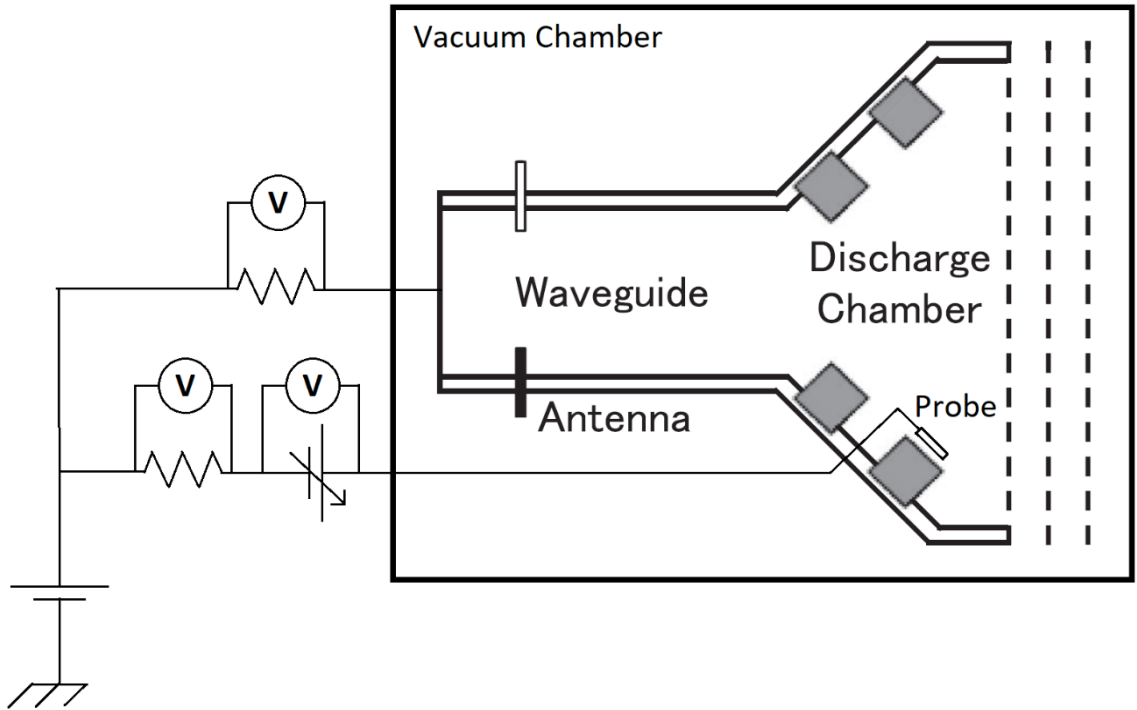

*Figure 7: Schematic of the Langmuir Probe Experiment Setup*

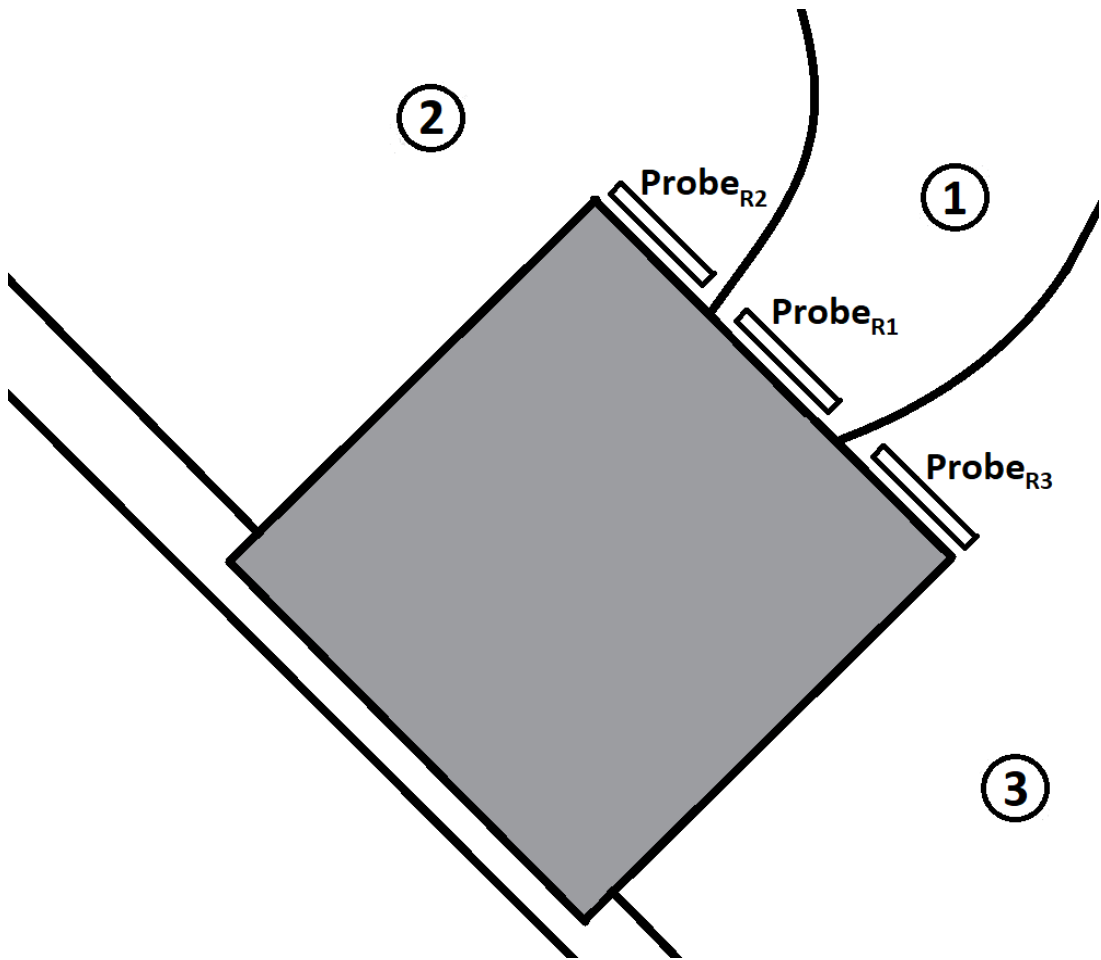

*Figure 8: Langmuir Probe Location in the Discharge Chamber (zoom on Fig. 7, downstream magnet)*

*Table 1: Plasma Properties in the Discharge Chamber Experiment Operating Conditions*

| Mass Flow | 2.4sccm |
|---|---|
| Input Power | 34W |
| Reflected Power | 0.4W |
| Screen Grid Voltage | 1500V |
| Acceleration Grid Voltage | -350V |
| Probe Voltage (A-P-L) | ±80V |
| Beam Current | 148mA |

A sample of the Langmuir probe raw data, collected under ion beam extraction, is shown in Figure 9. From the direct analysis of the Langmuir probe curves we obtain the values for $n_e$ and $T_{el}$ reported in Table 2.

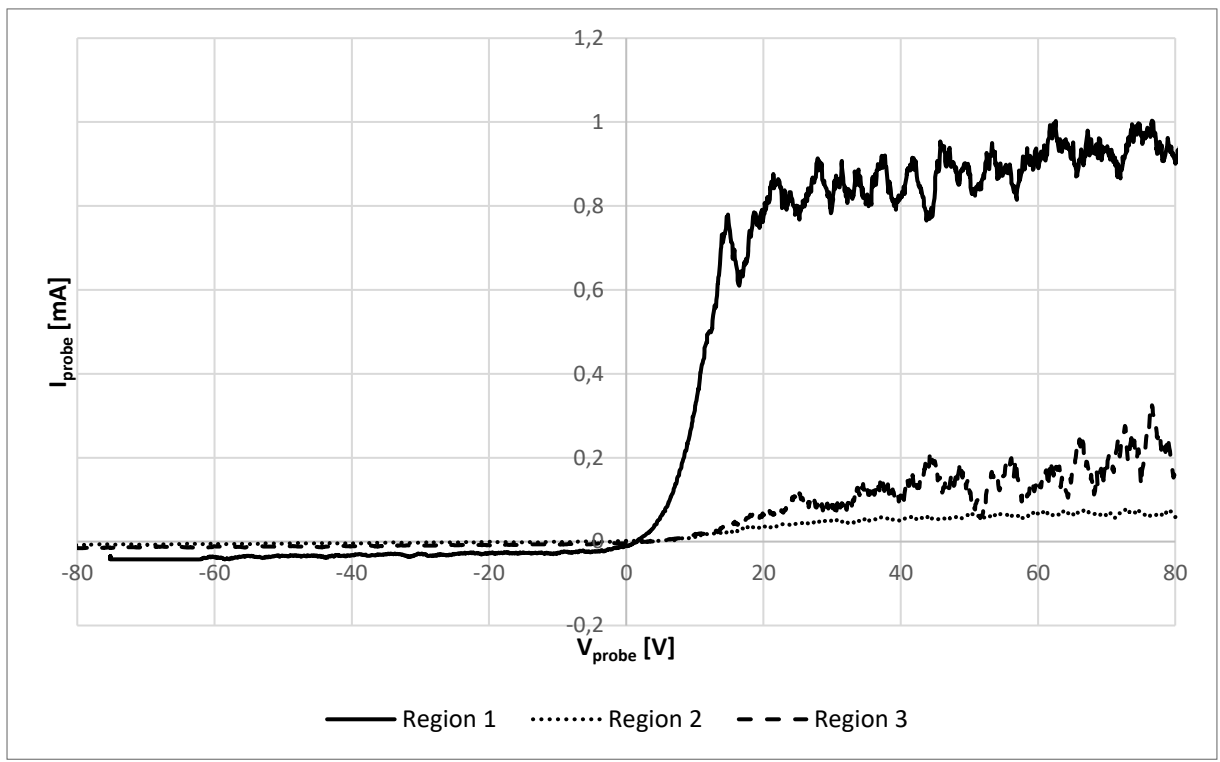

*Figure 9: Sample of the Langmuir Probe Data for the Three Regions*

*Table 2: Electron Density and Temperature in the Three Regions*

| | $n_e$ | $T_{el}$ |
|---|---|---|
| Region 1 | $3.3*10^{16}$ particles/m$^3$ | 4.0eV |
| Region 2 | $1.1*10^{15}$ particles/m$^3$ | 7.7eV |
| Region 3 | $4.4*10^{15}$ particles/m$^3$ | 7.8eV |

Density in Region 1 is one order of magnitude larger than Region 2 and 3, as predicted in section 2.2. On the other hand, an unexpected feature of these two is that they have a higher electron temperature compared to Region 1.

To obtain a complete analysis of the electron temperature distribution, we apply the so called "Medicus method" [11], which allows us to obtain it from the Langmuir probe curves. This is based on the equation:

$$F(V) = \left(\frac{4}{n_e}\right)\sqrt{\frac{m_e}{2\,e\,V}}\frac{\Delta I_{probe}}{\Delta V}$$

(15)

The resulting electron energy distribution, shown in Figure 10, show that the temperature distributions in Regions 2 and 3, while resembling the pattern of Region 1, are shifted towards higher temperatures. All these three are non-Maxwellian, a known feature of ECR discharges. The slightly higher temperature in Region 3 is plausibly due to the moderate ECR heating in the region.

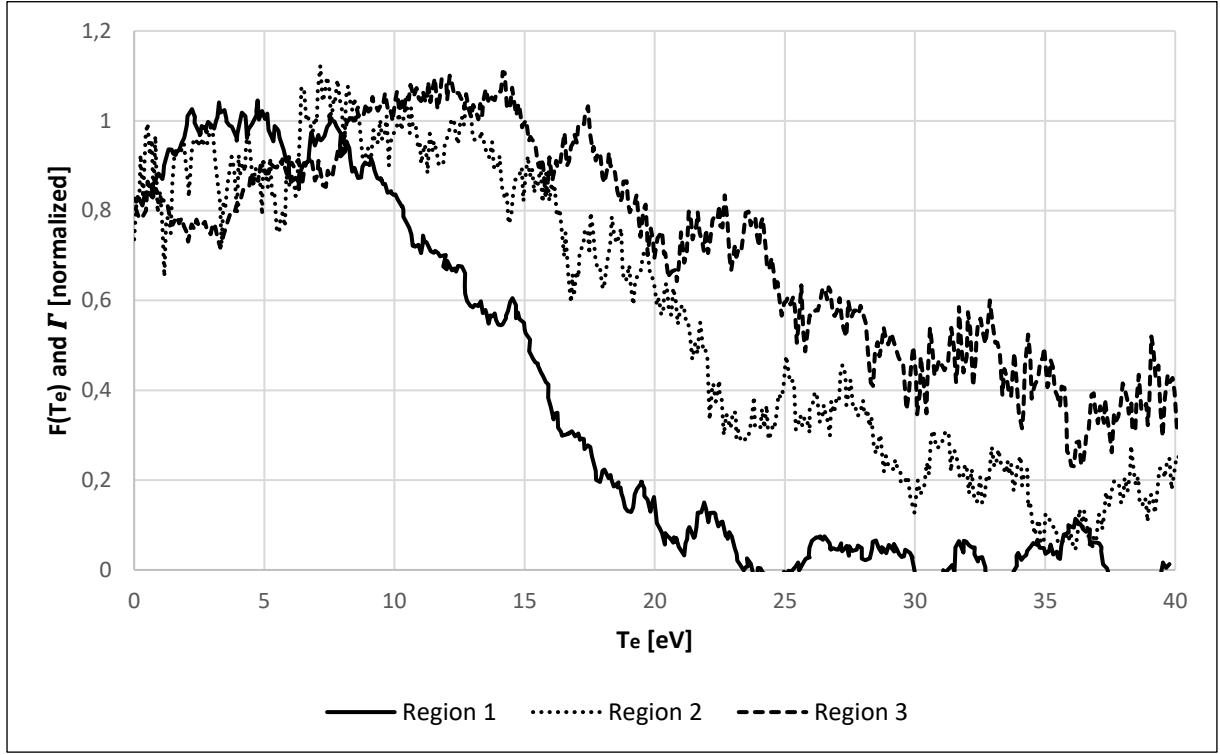

Figure 10: Electron Energy Distribution Function in the Three Regions

### 3.2. Measurement of the Microwave Power Absorption Efficiency by Curve Fitting

As mentioned in the methods section, we choose to measure the microwave power absorption efficiency with a non-invasive method as well as with Langmuir probes. The curve fitting approach we propose has a lower resolution compared to the Langmuir probe technique as it requires the assumption that α is independent on ṁ (its validity will be discussed in the next paragraph). Hence, we will obtain one value for each microwave power input.

In this experiment, µ10 is operated in a wider range of conditions, reported in Table 4, in order to visualize the different plasma modes occurrence. Errors within 3% might occur due to the beam current data resolution. The results will be shown by plotting $C_i$ as function of η, as from this plot the value of α is obtained by interpolating the left asymptote. This approach is visually represented in Figure 11.

Table 3: Microwave Power Absorption Efficiency Measurement by Curve Fitting Experiment Conditions

| Mass Flow | 1.35-3.3sccm (0.15sccm step) |
|---|---|
| Microwave Power | 28-40W (3W step) |
| Screen Grid Voltage | 1500V |

| Acceleration Grid Voltage | -350V |

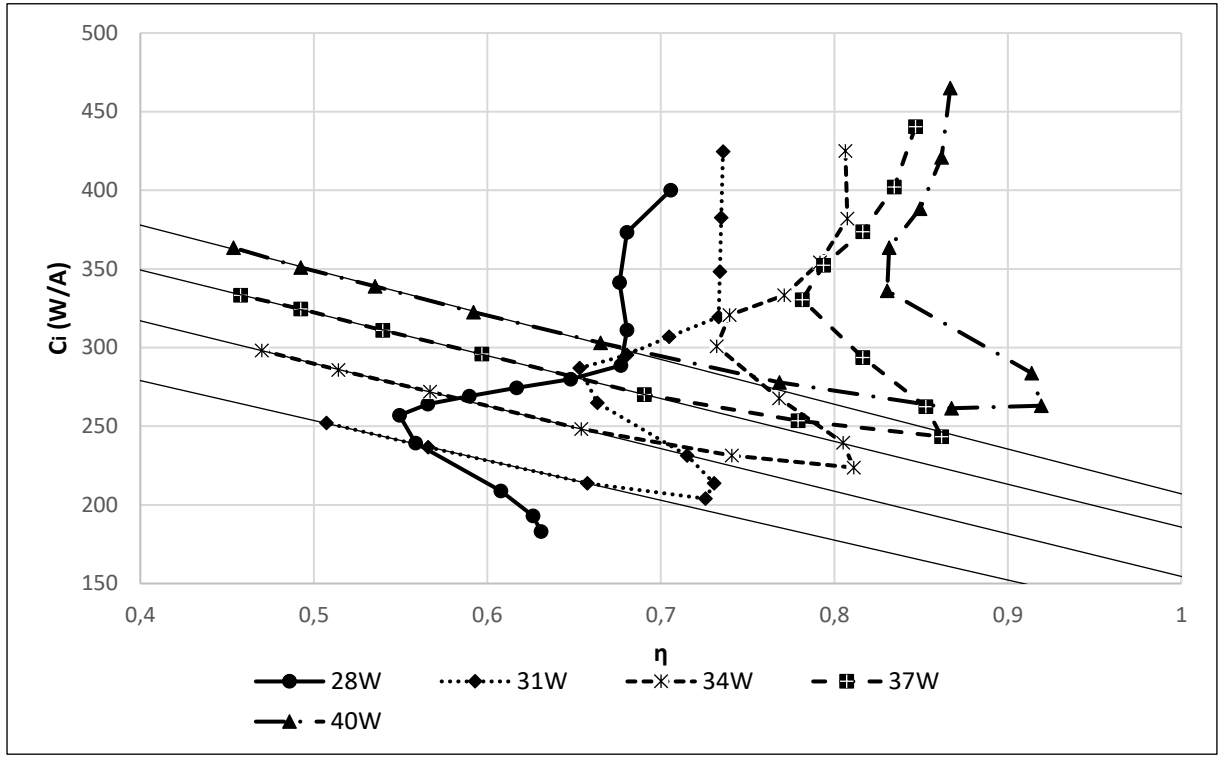

*Figure 11: Measurement of the Microwave Power Absorption Efficiency by Curve Fitting Approach*

We notice how, in agreement with the predictions of the 0-dimensional ion production model, the asymptote is not horizontal. As our assumption states that α is independent on ṁ (hence on η), we interpret this as a variation of $\varepsilon_p^*$ depending on η. To address this, we fit the asympthote linearly (with the parameters a and b) and assume that the optimal value of $\varepsilon_p^*$ (30eV) is reached at the optimal operation point, in agreement with the model [6], and obtain α as:

$$C_i = a - b\,\eta$$

(16)

$$\alpha = \frac{\varepsilon_{pOPT}^*}{C_{iOPT}\,f_s}$$

(17)

$C_{iOPT}$ is the ideal value of $C_i$ along the asymptote, calculated using the highest value of η in equation 16. We plot the results for α as function of $C_i$ in Figure 12: results are in good agreement with both the theory [6]. In this case, errors within 3% might occur due to the beam current data resolution.

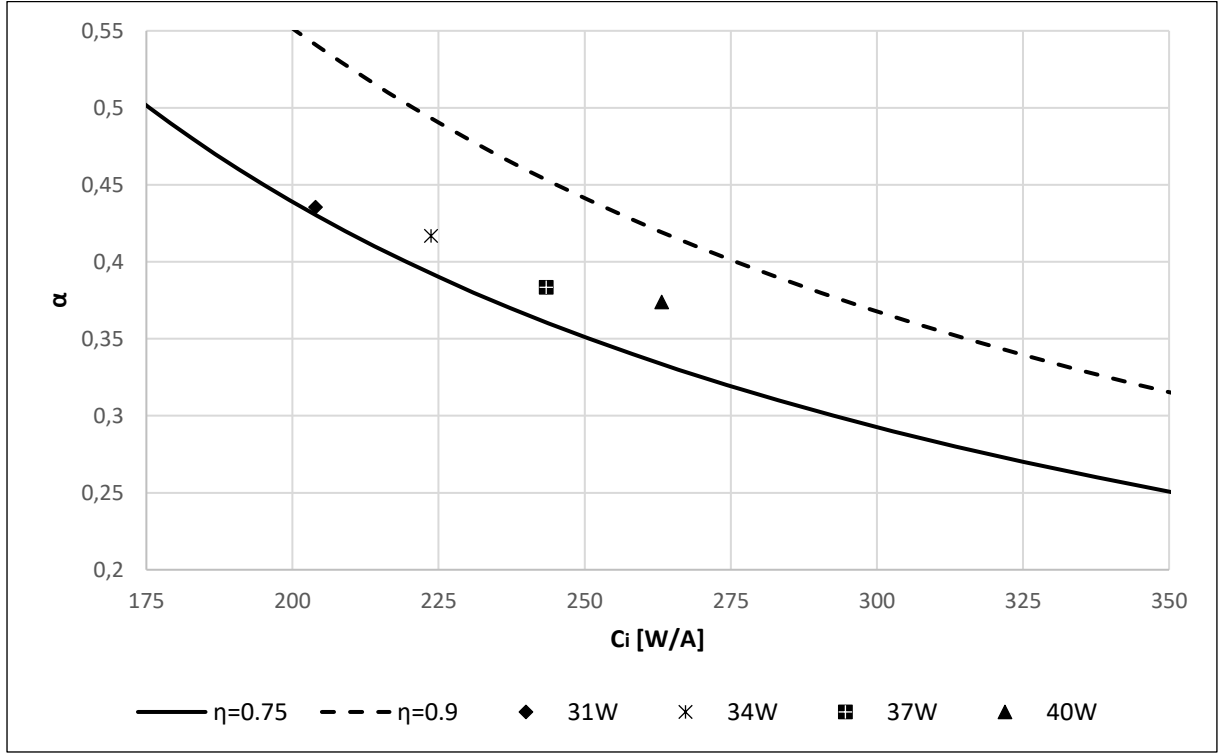

*Figure 12: Microwave Power Absorption Efficiency as Function of the Ion Production Cost (with Curve Fitting)*

### 3.3. Measurement of the Microwave Power Absorption Efficiency by Langmuir Probes

As pointed out by the three regions model, and confirmed by the largely different electron density measurements in 3.1, measuring the microwave power absorption efficiency is meaningful only in Region 1.

From a finite elements FEMM simulation of the magnetic field, we can obtain the geometrical parameters as: $A_{min}=27cm^2$, $\theta_{mag}=28°$, $\theta_{ECR}=54°$ and $\partial B/\partial s=0.1T/cm$. Data regarding $E_{ECR}$ is obtained from a previous investigation of the electric field intensity inside the discharge chamber as 1kV/m [10]. $n_e$ is measured with the central Langmuir probe (R1 probe in Figure 8). In this experiment, we measure α in multiple operating conditions, as reported in Table 3.

*Table 4: Microwave Power Absorption Efficiency Measurement by Langmuir Probes Experiment Conditions*

| Mass Flow | 2.1-2.85sccm (0.15sccm step) |
|---|---|
| Microwave Power | 28-40W (3W step) |
| Screen Grid Voltage | 1500V |
| Acceleration Grid Voltage | -350V |
| Probe Voltage (A) | ±80V |

We start by observing the results for α as function of $P_\mu$ (Figure 13) and $\dot{m}$ (Figure 14).

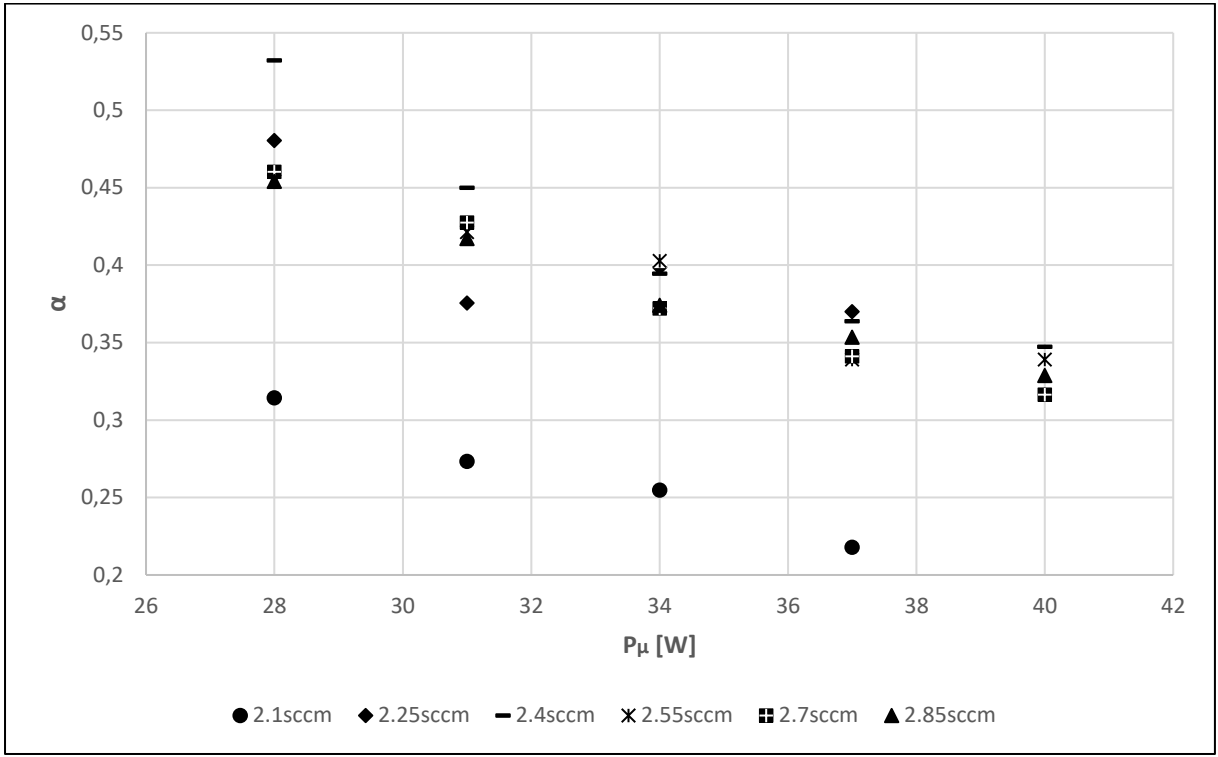

*Figure 13: Microwave Power Absorption Efficiency as Function of the Microwave Power*

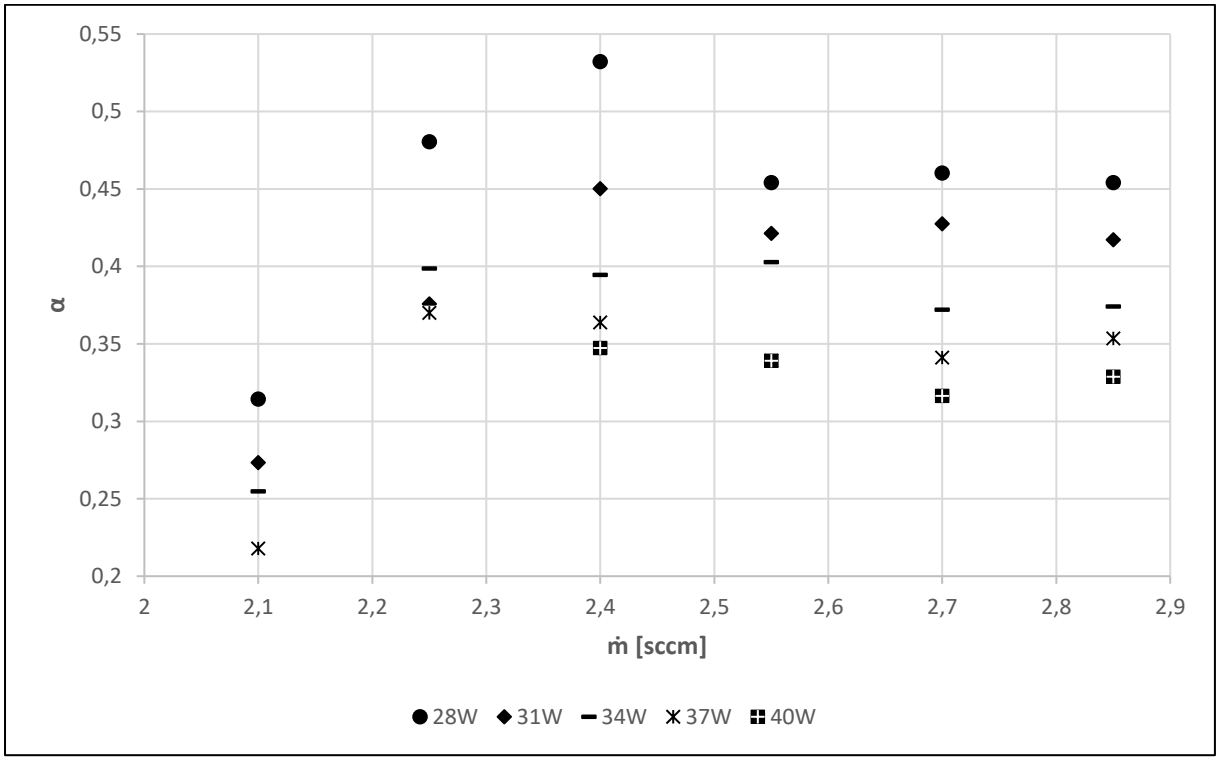

*Figure 14: Microwave Power Absorption Efficiency as Function of the Mass Flow*

We observe how, at 2.1sccm, the transition to high beam mode (more effective plasma generation) still has not occurred, which considerably lowers performance compared to higher mass flows. For all the other operating conditions, from the first graph we can observe a clear decreasing trend in α at higher $P_\mu$. On the other hand, no clear dependence is observable with ṁ, having small discrepancies that can be attributed to Langmuir probe disturbances and experimental error. Potential sources of error are the measured values of $E_{ECR}$ and $n_e$ (from the electron saturation current), within a range of 10%. The range of values observed for α has a good agreement with predictions coming from the 0-dimensional ion production model. We can observe this in Figure 15 by plotting the α as function of $C_i$ both for experimental values and theory predictions.

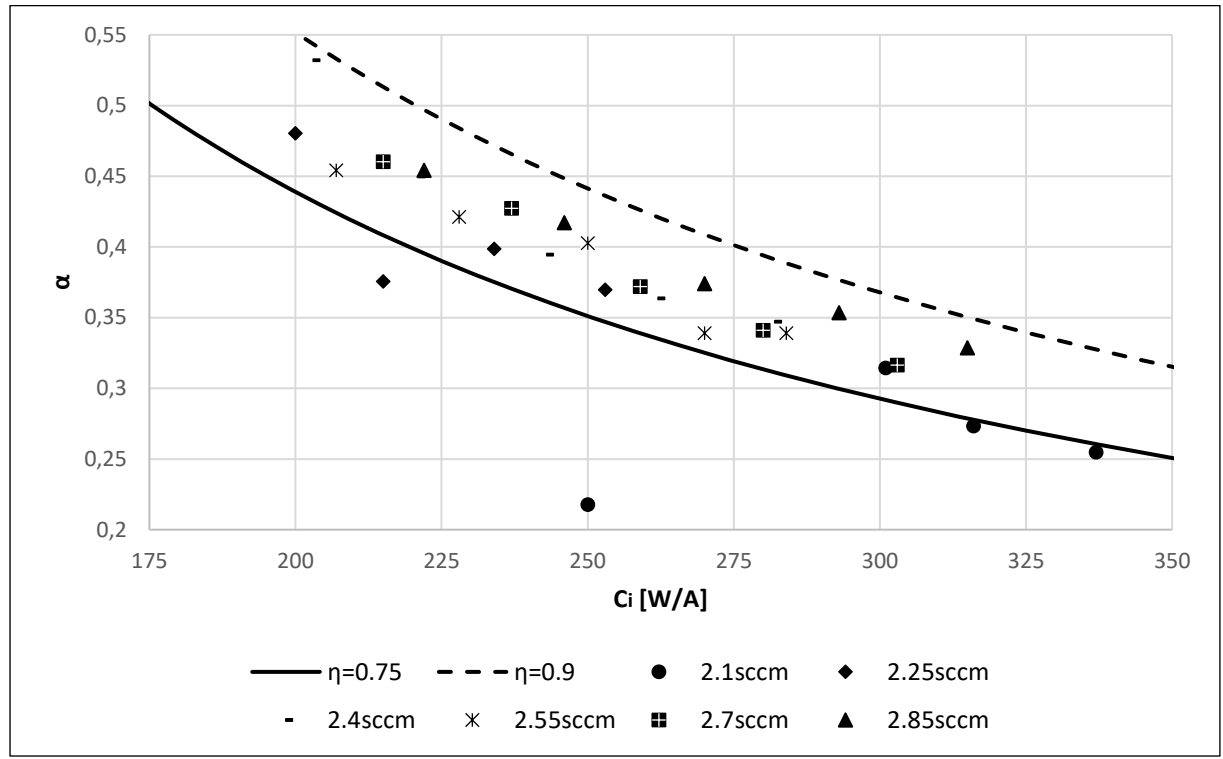

*Figure 15: Microwave Power Absorption Efficiency as Function of the Ion Production Cost (with Lanmguir Probes)*

### 4. Discussion
### 4.1. Interpretation of the Electron Temperature Results

The representation given in section 2.2, based on particle dynamics, would lead to presence of plasma only in the Region 1. However, due to the plasma collective behavior (diffusion), plasma is found also in the other two, as shown by the experiments in 3.1. Furthermore, the electron temperature distribution of Regions 2 and 3 considerably differs from what is observed in Region 1.

We try to give an interpretation of the phenomenon observed by introducing a multi-temperature diffusion model.

In diffusion theory, the particle flux is determined by [9]:

$$\boldsymbol{\Gamma} = \mathrm{n}\,\boldsymbol{v} = -D\,\boldsymbol{\nabla}n$$

(18)

Where D is the diffusion coefficient. A common assumption while analyzing ECR ion thrusters is weak ionization: charged particles will interact prevalently with neutrals [9]. Building upon this, we make the further consideration that no interaction with other charged particles implies no interaction with other charged particles with a different energy. This allows us to treat electrons with a different temperature as distinct species. Making this consideration, we can replace the density in equation 18 with an arbitrary electron distribution:

$$\boldsymbol{\Gamma}(v) = -D\, \boldsymbol{\nabla} F(v)$$

(19)

We show the implications of this approach for a Maxwell-Boltzmann distribution of electrons (peaking at 4eV, as in the case of µ10) in Region 1, and assuming $n_{R2}=n_{R3}=0$ in the other two, so that the gradient can be simplified (we have seen from the experiments how densities of Region 2 and 3 are small compared to Region 1). We disregard the actual value of $n_{act}$ and D by normalizing the results, in order to observe the general case. As we can see in Figure 16, this model predicts that the temperature of the electrons outflowing from Region 1 will have a peak at 10eV. The outcome will only depend on the F(v) and diffusion model selected (classical or neo-classical), while the magnetic field intensity will influence the actual value of Γ.

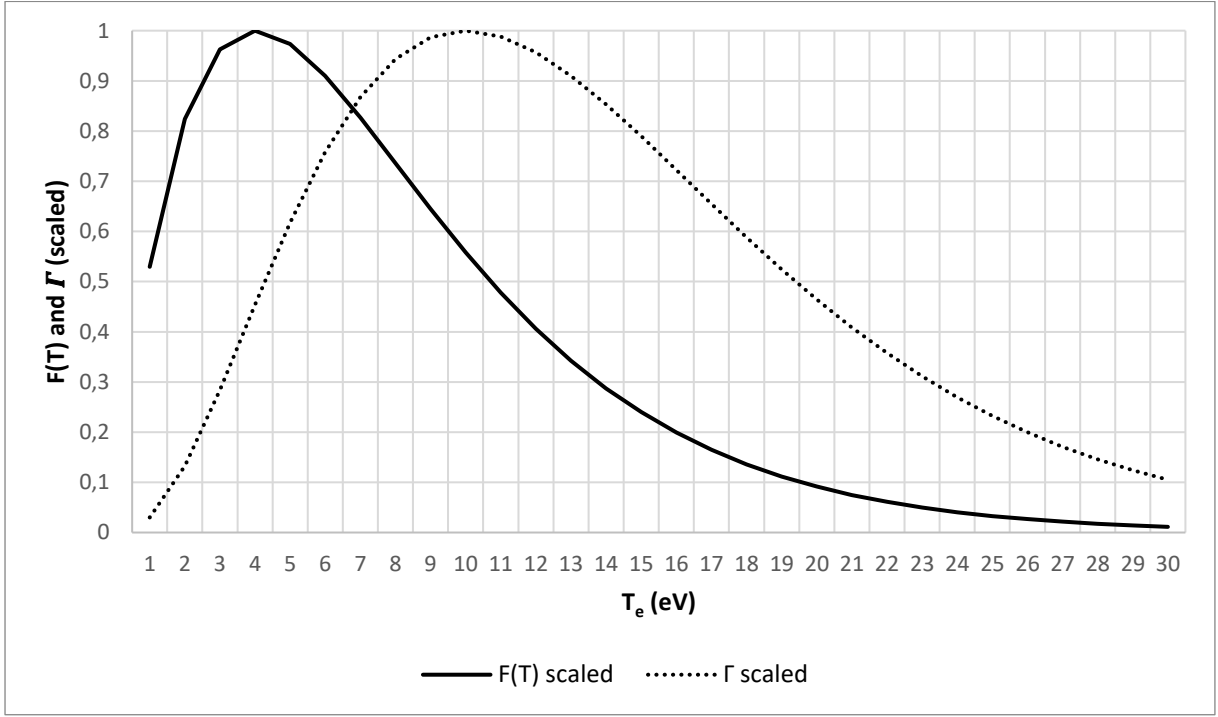

*Figure 16: 4eV Maxwellian Distribution and Predicted Outwards Flux*

Finally, we introduce the density and magnetic field parameters found in µ10 in the diffusion model we have proposed. Results are shown in Table 5.

|  | $\Gamma_{avg}$ |
|---|---|
| Region 1-Region 2 | $6.8*10^{13}$ particles/m² s |
| Region 1-Region 3 | $2.1*10^{14}$ particles/m² s |

*Table 5: Predicted Electron Flux from Region 1*

Based on this model, the average electron flux towards Region 3 is expected to be approximately three times the flux towards Region 2.

The diffusion model proposed brings theoretical justification, both in terms of electron energy and density, to the experiment results presented in 3.1.

### 4.2. Microwave Power Absorption Efficiency

This paper shows three methods to evaluate α. In the first, α is introduced from the point of view of the microwave power distribution in the discharge chamber. The second method is derived from the ion production characteristics. The third one is associated with the electron heating process. Investigation of the microwave power absorption efficiency added valuable information to the 0-dimensional ion production model for ECR discharge ion thruster previously developed, while also confirming the predicted values of 0.3-0.5. Based on these results, we can further calculate the bidirectional current at the ECR surface with equation 6: results range from 8A to 12A. The theoretical formulation of Eqs. 3 and 14 pointed out how α is affected by $B_{mag}$ and $B_{ECR}$: large $B_{mag}$ and small $B_{ECR}$ will improve α. $B_{mag}$ is dependent on the permanent magnet characteristic: since the space-qualified ion thrusters are operated at very high temperature, the Samarium-Cobalt magnets are utilized instead of Neodymium, so that $B_{mag}$ is around 3kG. $B_{ECR}$, on the other hand, is proportional to the microwave frequency: lowering it would potentially improve α, but due to the plasma cut-off phenomena this would lead to a lower plasma density, and hence thrust. The μ10 ion thruster utilizes 4GHz microwave as a tradeoff solution. On the other hand, $B_{min}$ has a relatively small influence on α, and can in fact be simplified for small values of $\theta_{min}$ and $\theta_{ECR}$.

Results from the Langmuir probe measurements and the curve fitting approach point out how α decreases for increasing microwave power input. On the other hand, the first also shows how this parameter is not affected by the mass flow. This cannot be verified with the curve fitting method, as it requires the independence on ṁ as an assumption. We can finally compare the results from the two different methods, as done in Fig. 17: averaging results for a certain $P_\mu$, Langmuir probes register regularly a 10% lower performance, compared to the curve fitting approach, both in terms of α and in terms of $C_i$. This can be easily explained by the disturbance caused by the Langmuir probes to the ECR discharge, which was observed during the experiments to decrease the beam current. Furthermore, the good agreement between these methods points out how power losses at the discharge chamber wall should be negligible.

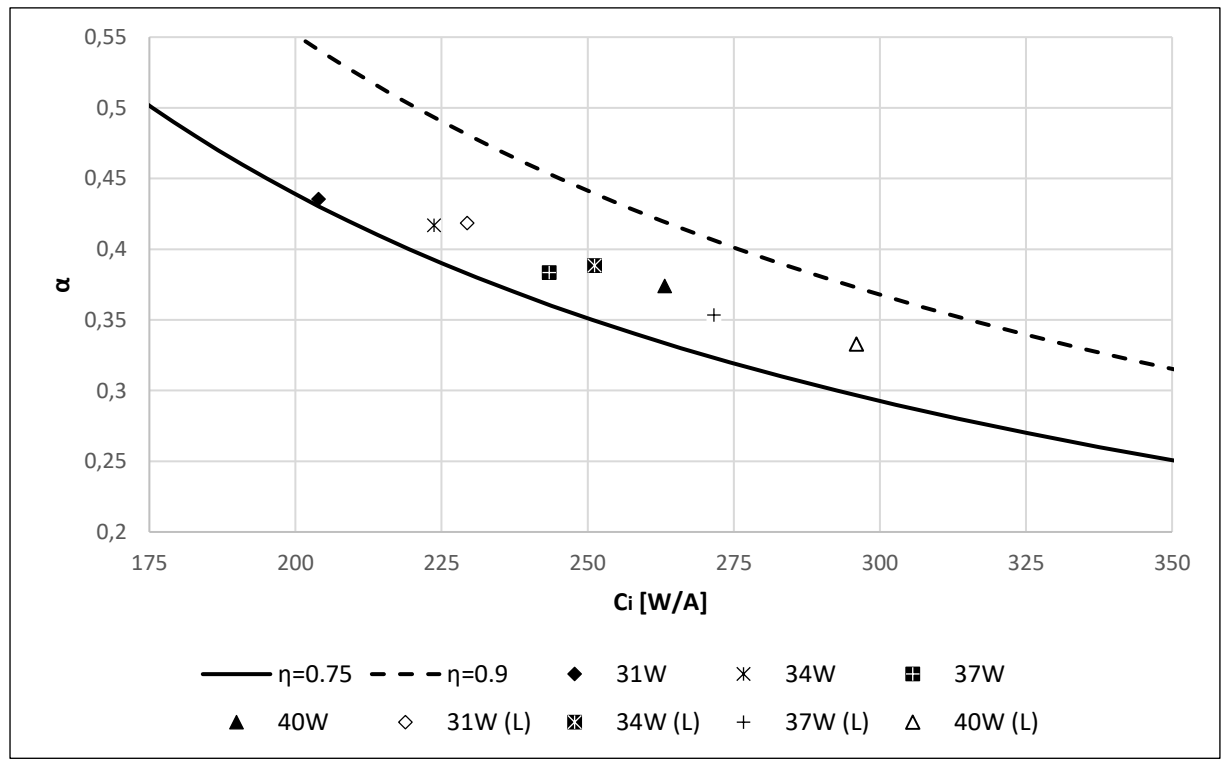

*Figure 17: Microwave Power Absorption Efficiency as Function of the Ion Production Cost (with Curve Fitting and Averaged Lanmguir Probes)*

Both the curve fitting and the Langmuir probe approaches to measure the microwave power absorption efficiency present sources of measurement uncertainty, that should be addressed in future research. In the first, we assume most parameters of equation 4 as constants to simplify the curve fitting. Further investigation on parameters such as $\varepsilon_p^*$ and $C_0$ and their dependence on $\dot{m}$ and $P_\mu$ would improve the accuracy of these measurements and further improve our understanding of the ECR ion thruster performance. As for the second method, other than the disturbance caused by the Langmuir probes themselves, the main uncertainty is about $E_{ECR}$. The electro-optical measurement technique utilized to measure this parameter [10] is complex to implement and has a relatively low resolution, so in this research we utilized a standard value of $E_{ECR}$ for the measurement of α. Improved measurement techniques for electric fields in ECR plasma would make this method more accurate.

## 5. Conclusion

The research presented in this paper focused on the modeling of the microwave power absorption to high energy electrons through ECR and its subsequent measurement. The results can be summarized as follows:

1) A three regions subdivision of the ECR discharge chamber was proposed and verified. Experimental results are consistent with the initial hypothesis that plasma ionization occurs virtually only in Region 1, where the magnetic mirror confine the high energy electrons gradually heated by passing multiple times through the ECR region. Furthermore, from these findings, a model attempting to explain the electron temperature distribution observed in Region 2 and 3 has been proposed.

2) An investigation of the microwave power absorption efficiency in the ECR ion thruster has been performed. It involved the derivation of equations useful to measure this parameter and to understand the physics behind the power absorption process. Three methods were proposed to measure it: an analysis of the microwave power distribution to the discharge chamber, a global measurement based on the ion production characteristics and a local measurement based on the electron heating process and performed with Langmuir probes. These measurements pointed out the dependence of the microwave power absorption efficiency with the power input, with values ranging from 0.3 to 0.5.

These conclusions provide a justification for previously achieved performance improvements [4]: as ionization occurs mostly in Region 1, injecting propellant from input ports close to it increases the thruster efficiency. Moreover, they are presently supporting the development of upgraded discharge chambers for the μ10 ion thruster. Further work in this field should involve a method to optimize the magnetic field design of the chamber to maximize the microwave power absorption efficiency by enlarging Region 1, increasing $B_{mag}$ or reducing $B_{ECR}$.